# Approximate Decoding Approaches for Network Coded Correlated Data

Hyunggon Park*, Nikolaos Thomos† and Pascal Frossard†

*Multimedia Communications and Networking Laboratory, Ewha Womans University, Seoul, Korea.

†Signal Processing Lab. (LTS4), Ecole Polytechnique Fédérale de Lausanne (EPFL), Lausanne, Switzerland

hyunggon.park@ewha.ac.kr, {nikolaos.thomos, pascal.frossard}@epfl.ch

**Abstract**

This paper considers a framework where data from correlated sources are transmitted with help of network coding in ad-hoc network topologies. The correlated data are encoded independently at sensors and network coding is employed in the intermediate nodes in order to improve the data delivery performance. In such settings, we focus on the problem of reconstructing the sources at decoder when perfect decoding is not possible due to losses or bandwidth bottlenecks. We first show that the source data similarity can be used at decoder to permit decoding based on a novel and simple *approximate decoding* scheme. We analyze the influence of the network coding parameters and in particular the size of finite coding fields on the decoding performance. We further determine the optimal field size that maximizes the expected decoding performance as a trade-off between information loss incurred by limiting the resolution of the source data and the error probability in the reconstructed data. Moreover, we show that the performance of the approximate decoding improves when the accuracy of the source model increases even with simple approximate decoding techniques. We provide illustrative examples about the possible of our algorithms that can be deployed in sensor networks and distributed imaging applications. In both cases, the experimental results confirm the validity of our analysis and demonstrate the benefits of our low complexity solution for delivery of correlated data sources.

This work has been supported by the Swiss National Science Foundation (grants PZ00P2-121906 and 200021-118230) and by Basic Science Research Program through the National Research Foundation of Korea (NRF) funded by the Ministry of Education, Science and Technology (2010-0009717). This work was mostly performed while the first author was with EPFL.

# I. Introduction

The rapid deployment of distributed networks such as sensor networks, cloud networks has motivated a plethora of researches that study the design of low complexity and efficient solutions for information delivery. Since the coordination among intermediate nodes is often difficult to achieve, the information dissemination in the intermediate nodes has often to be performed in a distributed manner on ad-hoc or overlay mesh network topologies. Network coding [1] has been recently proposed as a method to build efficient distributed delivery algorithms in networks with path and source diversity. It is based on a paradigm where the network nodes are allowed to perform basic processing operations on information streams. The network nodes can combine information packets and transmit the resulting data to the next network nodes. Such a strategy permits to improve the throughput of the system and to approach better the max-flow min-cut limit of networks [2], [3]. When the decoder receives enough data, it can recover the original source information by performing inverse operations (e.g., with Gaussian elimination).

These advantages motivate the deployment of network coding in various scenarios where the network diversity is significant (e.g., [4]–[9]). Many of these solutions are based on random linear network coding (RLNC) [10] that permits to implement distributed solutions with low communication costs. RLNC represents an interesting solution for the deployment of practical systems where it can work in conjunction with data dissemination protocols such as gossiping algorithms [8]. The resulting systems are robust against link failures, do not require reconciliation between the network node, and can significantly improve the performance of data delivery compared to ’store and forward’ approaches. Most of research so far has however focused either on theoretical aspects of network coding such as achievable capacity and coding gain, or on its practical aspects such as robustness and increases throughput when the number of innovative packets is sufficient for perfect decoding. However, it generally does not consider the problematic cases where the clients receive an insufficient number of innovative packets for perfect decoding due to losses or timing constraints for example. This is the main problem addressed in this paper.

We consider a framework where network coding is used for the delivery of correlated data that are discretized and independently encoded at the sensors. The information streams are delivered with help of RLNC in lossy ad-hoc networks. When an insufficient number of symbols at decoder prevent exact data recovery, we design a novel low complexity *approximate decoding* algorithm that uses the data correlation for signal reconstruction. The information about source similarity typically provides additional constraints in the decoding process, such that well-known approaches for matrix inversion (e.g., Gaussian elimination) can be efficiently used even in the case where the decoding problem is a priori underdetermined. We show analytically that the use of source models at decoding process leads to an improved data recovery. Then, we analyze the impact of accurate knowledge of data similarity at decoder, where more precise information leads to better performance in the approximate decoding. We further analyze the influence of the choice of the Galois Field (GF) size in the coding operations on the performance of the approximate decoding framework. We demonstrate that the field size should be selected by considering the tradeoff between resolution in representing the source and approximate decoding performance. Specifically, when the GF size increases, the quantization error of the source data decreases while the decoding error probability increases with the GF size. We show that there is an optimal value for the GF size when the approximate decoding is enabled at the receivers. Finally, we illustrate the performance of the network coding algorithm with the approximate decoding on two types of correlated data, i.e., seismic data and video sequences. The simulation results confirm the validity of the GF size analysis and show that the approximate decoding scheme leads to efficient reconstruction when the accurate correlation information is used during decoding. In summary, the main contributions of our paper are (i) a new framework for the distributed delivery of correlated data with network coding, (ii) a novel approximate decoding strategy that exploits the data similarity with low complexity when the received data does not permit perfect decoding, (iii) an analysis of the influence of the accuracy of the data similarity information and the GF size on the decoding performance, and (iv) the implementation of illustrative examples with external or intrinsic source correlation.

In general, the transmission of correlated sources is studied in the framework of distributed coding [11] (i.e., in the context of Slepian-Wolf problem), where sources are typically encoded by systematic channel encoders and eventually decoded jointly [12], [13]. DSC (distributed source coding) is also combined with network coding schemes [14]–[17] in the gathering of correlated data. Our focus is however not on the design of a distributed compression scheme, which generally assumes that sensors are aware of the similarity between the data sources. Rather, we focus on the transmission of correlated data that are encoded independently, transmitted with help of network coding over an overlay network and jointly decoded at the receivers. However, due to the network dynamics, there is no guarantee that each node receives enough useful packets for successful data recovery. Hence, it is essential to have a low complexity methodology that enables the recovery of the original data with a good accuracy, when the number of useful packets is not sufficient for perfect decoding. When RLNC is implemented in the network, the encoding and decoding processes of each node are based on linear operations (e.g., linear combinations, inverse of linear matrix, etc.) in a finite algebraic field. In the case of insufficient number of innovative packets for perfect decoding, one can simply deploy an existing regularization technique that may minimize the norm of the errors using the pseudo-inverse of the encoding matrix. However, it is generally known that this type of regularization techniques may result in significantly unreasonable approximation [18]. Alternatively, Tikhonov regularization provides an improved performance by slightly modifying the standard least square formula. However, this technique requires to determine additional optimization parameters, which is nontrivial in practice. Sparsity assumptions might also be used [19] for regularized decoding in underdetermined systems in cases where a sparse model of the signal of interest is known a priori. However, all these regularization techniques have been designed and developed in the continuous domain, but not for finite fields that are used in network coding approaches. Thus, they may show significantly poor performance if they are blindly applied in our framework, as they cannot consider several properties (e.g., cyclic) of finite field operations. Underdetermined systems can also be solved approximately based on the maximum likelihood estimation (MLE) techniques (see e.g., [20] (Part II)), but these techniques require effective data models and typically involve large computational complexity.

The paper is organized as follows. In Section II, we present our framework and describe the approximate decoding algorithm. We discuss the influence of the source model information in the approximate decoding process in Section III. In Section IV, we analyze the relation between the decoding performance and the GF size, and then determine an optimal GF size that achieves the smallest expected decoding error. Section V and Section VI provide illustrative examples that show how the proposed approach can be implemented in sensor network or video delivery applications.

## II. Approximate Decoding Framework

We begin to describe the general framework considered in this paper and present the proposed distributed delivery strategy for correlated data sources. We also discuss the concept of approximate decoding that enables receivers to estimate the source information when the number of data packets is not sufficient for perfect decoding.

### A. RLNC Encoding

We consider an overlay network with sources, intermediate nodes, and clients distributed over a network (e.g., ad-hoc network). We denote by $s_1, \ldots, s_N$ the symbols generated by $N$ discrete and correlated sources, where $s_n \in \mathcal{S}(\subset \mathbb{R})$ for $1 \leq n \leq N$. $\mathcal{S}$ is an alphabet set of $s_n$ and $|\mathcal{S}|$ denotes the size of $\mathcal{S}$. These source data are transmitted to the clients via intermediate nodes that are able to perform network coding (i.e., RLNC). Hence, each $s_n$ also needs to be considered as an element in a GF. In order to explicitly specify whether $s_n$ is in the field of real numbers or in a GF, we define *identity functions*, defined as

$$\begin{cases} 1_{\mathbb{R}G} : \mathbb{R} \to GF, & 1_{\mathbb{R}G}(s_i) = x_i \\ 1_{G\mathbb{R}} : GF \to \mathbb{R}, & 1_{G\mathbb{R}}(x_i) = s_i \end{cases} \tag{1}$$

which means that $x_i$ is an element in GF representing $s_i$. Thus, an intermediate node $k$ using RLNC transmits a packet generated as

$$y(k) = \sum_{n=1}^{N} \bigoplus \{c_n(k) \otimes x_n\} \triangleq (c_1(k) \otimes x_1) \oplus (c_2(k) \otimes x_2) \oplus \cdots \oplus (c_N(k) \otimes x_N)$$

which is a linear combination of $x_n$ and coding coefficients $c_n(k)$ in GF. $\oplus$ and $\otimes$ denote an additive operation and a multiplicative operation defined in GF, respectively. The coding coefficients are uniformly and randomly chosen from GF with size $2^r$, denoted by $GF(2^r)$. This implies that the GF size is determined by $r$ and that $c_n(k) \in GF(2^r)$. In our implementation, the addition in GF with characteristic 2, i.e., $GF(2^r)$, is performed by the exclusive-OR (XOR) operation. The size of the field determines the set of coding operations that can be performed on source symbols. We thus assume that the size of the input set is $|\mathcal{S}| \leq 2^r$. If $|\mathcal{S}| > 2^r$, the input set is reduced (using e.g., source binning or quantization), such that the input set does not exceed the GF size (i.e., $2^r$).

The encoded symbols in each node are transmitted to neighboring nodes towards the client nodes. If a decoder receives $K$ innovative (i.e., linearly independent) symbols $y(1), \ldots, y(K)$, where all $y(k) \in GF(2^r)$, a linear system $\mathbf{y} = \mathbf{C} \odot \mathbf{x}$ can be formed as[1]

$$\begin{bmatrix} y(1) \\ \vdots \\ y(K) \end{bmatrix} = \begin{bmatrix} \mathbf{c}_1 & \cdots & \mathbf{c}_N \end{bmatrix} \odot \begin{bmatrix} x_1 \\ \vdots \\ x_N \end{bmatrix} \triangleq \sum_{n=1}^{N} \bigoplus \{\mathbf{c}_n \otimes x_n\} \tag{2}$$

where $\odot$ denotes the multiplication between matrices in a finite field. The $K \times N$ matrix $\mathbf{C}$ is referred to as the coding coefficient matrix, which consists of column vectors $\mathbf{c}_n = [c_n(1), c_n(2), \cdots, c_n(K)]^T$, where $\mathbf{A}^T$ denotes the transpose of a matrix $\mathbf{A}$. An illustrative example for $N = 3$ is shown in Fig. 1, where the symbols $s_1$, $s_2$, and $s_3$, which are mapped into $x_1$, $x_2$ and $x_3$ respectively, from sources are network encoded at intermediate nodes using randomly chosen coding coefficients.

### *B. Approximate Decoding*

Upon receiving a set of symbols $\mathbf{y}$ generated by (2), the decoder attempts to recover the source data. If $K = N$, i.e., the coding coefficient matrix $\mathbf{C}$ is full-rank as $N$ innovative symbols are available, then $\mathbf{x}$ is uniquely determined as $\mathbf{x} = \mathbf{C}^{-1} \odot \mathbf{y}$ (and correspondingly, $\mathbf{s} = 1_{G\mathbb{R}}(\mathbf{x})$) from the linear system in (2). Note that $\mathbf{C}^{-1}$ represents the inverse of the coding coefficient matrix $\mathbf{C}$ and can be obtained by well-known approaches such as the Gaussian elimination method over a GF.

However, if the number of received symbols is insufficient (i.e., $K < N$), there may be an infinite number of solutions $\hat{\mathbf{x}} = [\hat{x}_1, \ldots, \hat{x}_N]^T$ to the system in (2), as $\mathbf{C}$ is not full-rank. Hence, additional constraints should be imposed so that the coding coefficient matrix becomes full-rank. Hence, we modify the decoding system in (2) in order to include external information as coding constraints that permits decoding. This leads to approximate decoding, where the correlation of the input data is exploited to construct additional constraints $\mathbf{D}$ (all elements of $\mathbf{D}$ are in GF as well) and $\boldsymbol{\nu}$ in the decoding process so that the system becomes solvable. With the additional constraints, $\mathbf{D}$ and $\boldsymbol{\nu}$, an approximate decoding solution can be expressed as

$$\hat{\mathbf{x}} = \begin{bmatrix} \mathbf{C} \\ \mathbf{D} \end{bmatrix}^{-1} \odot \begin{bmatrix} \mathbf{y} \\ \boldsymbol{\nu} \end{bmatrix} \tag{3}$$

which again can be implemented by the Gaussian elimination method in a finite field. The additional constraints $\mathbf{D}$ and $\boldsymbol{\nu}$ typically depend on the problems under consideration, i.e., the source models.[2]

[1] In this paper, vectors and matrices are represented by boldfaced lowercase and boldfaced capital letters, respectively.

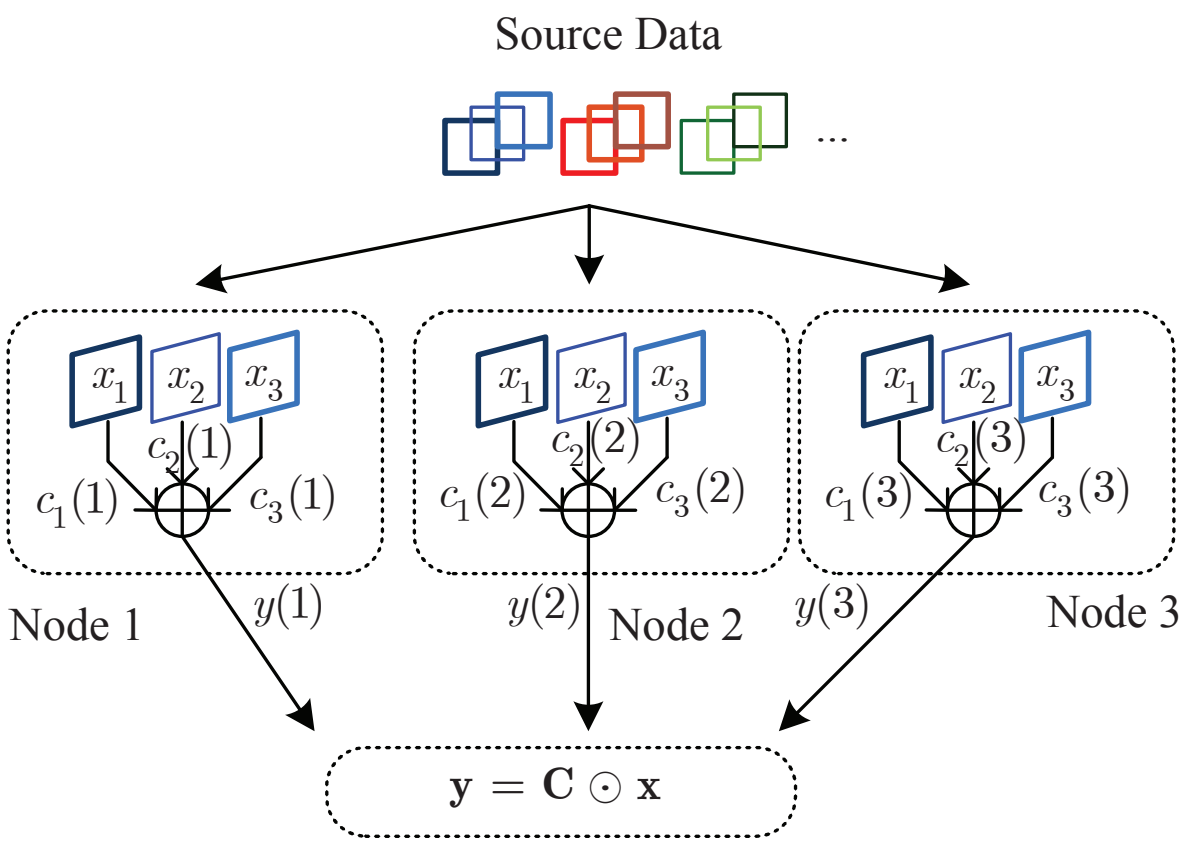

Fig. 1. Illustrative example of network coding with $N = 3$ source data and three network coding nodes. The input data $s_n$, which is mapped into $x_n$ in GF, are linearly combined with random coefficients in each network coding node, to generate vector $\mathbf{y}$.

An approximation $\hat{\mathbf{s}}$ of the original data can then be obtained by the identity functions defined in (1), i.e., $\hat{\mathbf{s}} = 1_{G\mathbb{R}}(\hat{\mathbf{x}})$. The distortion between $\mathbf{s}$ and $\hat{\mathbf{s}}$ is denoted by $\|\mathbf{s} - \hat{\mathbf{s}}\|_l$, where $\|\cdot\|_l$ denotes the $l-$norm operation [21]. An illustrative example of approximate decoding algorithm is described in Algorithm 1.

### C. *Simple Implementation of Approximate Decoding*

While the approximate decoding framework is generic, we present a simple instance of the algorithm in this paper.[3] Thus, our focus is on highlighting the potential advantages achieved by deploying simple approximate decoding approach for delivery of correlated data in resource constrained environments. Since $K$ innovative symbols are received, the rank of $\mathbf{C}$ in (3) is $K$, and correspondingly, $\mathbf{D}$ (3) is a $(N-K) \times N$ matrix of coefficients. The coefficients in $\mathbf{D}$ are determined based on the source correlation or similarity model. The source similarity is measured by the distance between data [22], [23]. More

[2] Alternatively, the source model information and the received symbols can be translated from GF into the field of real numbers, and the decoding process is performed. However, this may incur more computational complexity.

[3] By deploying more general source models and sophisticated algorithms on top of the proposed framework, better performance can be achieved.

**Algorithm 1** Approximate Decoding

**Given:** received symbols $\mathbf{y}$, coefficient matrix $\mathbf{C}$, data source model, data size $N$, GF size $2^r$.

1: **if** $rank(\mathbf{C}) = N$, **then**

2: $\quad \hat{\mathbf{s}} = 1_{G\mathbb{R}} \left( \mathbf{C}^{-1} \odot \mathbf{y} \right)$

3: **else** // $rank(\mathbf{C}) < N$ `and use approximate decoding`

4: $\quad$ Construct $\mathbf{D}$ and $\boldsymbol{\nu}$ based on available source model information

5: $\quad$ Compute $\hat{\mathbf{s}} = 1_{G\mathbb{R}} \left( \begin{bmatrix} \mathbf{C} \\ \mathbf{D} \end{bmatrix}^{-1} \odot \begin{bmatrix} \mathbf{y} \\ \boldsymbol{\nu} \end{bmatrix} \right)$

6: **end if**

specifically, the most similar data $s_i$ and $s_j$ have the smallest distance $|s_i - s_j|$. Then, we construct $\mathbf{D}$ with each row consisting of zeros, (i.e., additive identity of GF($2^r$)), except two elements of value "1" and "1" (because 1 is also an additive inverse of 1 in GF($2^r$)) that correspond to the positions of the most similar data $x_i$ and $x_j$. Accordingly, $\boldsymbol{\nu}$ is set as a zero vector with size of $(N - K)$, which is also appended to $\mathbf{y}$ and represent the results of the additional conditions set in $\mathbf{D}$. Thus, the implementation is expressed as

$$\hat{\mathbf{x}} = \begin{bmatrix} \mathbf{C} \\ \mathbf{D} \end{bmatrix}^{-1} \odot \begin{bmatrix} \mathbf{y} \\ \mathbf{0}_{(N-K)} \end{bmatrix}. \tag{4}$$

This enables the decoder to reconstruct the original symbols whenever the number of symbols is not sufficient for perfect decoding. With these additional equations, the decoder can then invert the linear system and approximate the data $\mathbf{x}$ with classical decoding algorithms.

Note that the coding coefficient matrix in (4) is assumed here to be non-singular, which happens with high probability if the size of the GF is large enough. However, the probability that the coding coefficient matrix becomes singular increases as the size of $\mathbf{D}$ is enlarged. In this case, the system includes a large number of similarity-driven coefficient rows with respect to the random coefficients of the original coding matrix. The impact of the singularity of the coding coefficient matrix on the performance of the approximate decoding is quantified in Section VI-B. Finally, we generally consider that there exists a solution to the decoding problem formed by the augmented coefficient matrix in (4). Otherwise, the decoder outputs a decoding error signal.

We study in the next sections the influence of the accuracy of the source model information and the influence of the finite field size (GF size) in the proposed approximate decoding algorithm. Specific implementations of the approximate decoding are later discussed in detail in Section V and Section VI with illustrative examples.

## III. Approximate Decoding based on A Priori Information on Source Model

We discuss in this section the performance of the proposed approximate decoding algorithm for recovering the source data from an insufficient number of network coded packets. In particular, we analyze and quantify the impact of the accuracy of the source model information (i.e., expected similarity between source values) at decoder when the augmented system in (4) enforces that the most similar data should have similar values after decoding. Recall that if approximate decoding is not deployed, conventional network decoding approaches for the network coded data cannot recover any source data.

We first show that the decoding error in our approximate decoding algorithm can be upper bounded as source data are more similar. This is described in Property 1.

*Property 1: The reconstruction error decreases as the sources are more similar.*

*Proof:* Let $\mathbf{y}$ be a set of $K$ received innovative packets (with $K$ smaller than the number of original symbols $N$, i.e., $K < N$). Let further $\mathbf{C}$ be the corresponding coding coefficient matrix and $\mathbf{x}$ be original source data as in (2). Since only $K < N$ innovative packets are available at decoder, $(N-K)$ additional constraints are imposed into the coding coefficient matrix $\mathbf{D}$ based on the approach discussed in Section II-C. This leads to the approximate decoding solution $\hat{\mathbf{x}}$ in (4).

We now analyze the error incurred by the proposed approximate decoding algorithm. The recovered symbol $\hat{\mathbf{s}} = 1_{G\mathbb{R}}(\hat{\mathbf{x}})$ from the approximate solution $\hat{\mathbf{x}}$ is compared to the exact solution $\mathbf{s}$. This exact solution is reconstructed based on the set of coding coefficients $\mathbf{C}$ and the coefficients $\mathbf{D}$, but with the exact constraints $\mathbf{d}$ (all the elements in $\mathbf{d}$ are in $GF(2^r)$) and not their approximation by a zero vector as done in (4). We denote these actual constraints by the vector $\mathbf{d}$, defined as

$$\mathbf{d} = \mathbf{D} \odot \mathbf{x} = [d(1), \ldots, d(N-K)]^T \tag{5}$$

which is computed by applying the additional coefficients in $\mathbf{D}$ on the original vector $\mathbf{x}$. Equivalently, $\mathbf{x}$ can be computed by

$$\mathbf{x} = \begin{bmatrix} \mathbf{C} \\ \mathbf{D} \end{bmatrix}^{-1} \odot \begin{bmatrix} \mathbf{y} \\ \mathbf{d} \end{bmatrix}. \tag{6}$$

Note that $\hat{\mathbf{x}}$ in (4) and $\mathbf{x}$ in (6) are obtained based on the operations defined in $GF(2^r)$, and thus, the resulting elements in $\mathbf{x}$ or $\hat{\mathbf{x}}$ are in $GF(2^r)$. However, they originally represent data in $\mathbb{R}$ (e.g., source data). Hence, in order to quantify the performance of the proposed algorithm, we are interested in the error between the exact and approximate solutions, i.e., $\|\mathbf{s} - \hat{\mathbf{s}}\|_l$.

From the assumption that $\left[\mathbf{C}^T \ \mathbf{D}^T\right]^T$ in (4) is not singular, its inverse, $\left[\mathbf{C}^T \ \mathbf{D}^T\right]^{-T}$ can be written as $\left[M_{(K)} \ M_{(N-K)}\right] = \left[\mathbf{m}_{(1)} \ \cdots \ \mathbf{m}_{(K)} \ \mathbf{m}_{(K+1)} \ \cdots \ \mathbf{m}_{(N)}\right]$, where $M_{(K)}$ and $M_{(N-K)}$ indicate sub-matrices with $\{\mathbf{m}_{(1)}, \ldots, \mathbf{m}_{(K)}\}$ and $\{\mathbf{m}_{(K+1)}, \ldots, \mathbf{m}_{(N)}\}$ column vectors. Thus, $\hat{\mathbf{s}}$ and $\mathbf{s}$ can be expressed from (4) and (6), respectively, as

$$\hat{\mathbf{s}} = 1_{G\mathbb{R}}(\hat{\mathbf{x}}) = 1_{G\mathbb{R}}\left(M_{(K)} \odot \mathbf{y}\right) \tag{7}$$

$$\mathbf{s} = 1_{G\mathbb{R}}(\mathbf{x}) = 1_{G\mathbb{R}}\left((M_{(K)} \odot \mathbf{y}) \oplus (M_{(N-K)} \odot \mathbf{d})\right). \tag{8}$$

Therefore, the error between the exact and the approximate solutions can be expressed as

$$\begin{aligned}
\|\mathbf{s} - \hat{\mathbf{s}}\|_l &= \|1_{G\mathbb{R}}(\mathbf{x}) - 1_{G\mathbb{R}}(\hat{\mathbf{x}})\|_l && (9)\\
&= \left\|1_{G\mathbb{R}}\left\{(M_{(K)} \odot \mathbf{y}) \oplus (M_{(N-K)} \odot \mathbf{d})\right) - 1_{G\mathbb{R}}\left(M_{(K)} \odot \mathbf{y}\right\}\right\|_l && (10)\\
&\le \left\|1_{G\mathbb{R}}\left\{(M_{(K)} \odot \mathbf{y}) \oplus (M_{(N-K)} \odot \mathbf{d}) \oplus (M_{(K)} \odot \mathbf{y})\right\}\right\|_l && (11)\\
&= \left\|1_{G\mathbb{R}}\left(M_{(N-K)} \odot \mathbf{d}\right)\right\|_l \\
&= \left\|1_{G\mathbb{R}}\left(\sum_{k=1}^{N-K} \bigoplus \{\mathbf{m}_{K+k} \otimes (x_{i,k} \oplus x_{j,k})\}\right)\right\|_l && (12)\\
&\le \left\|\sum_{k=1}^{N-K} 1_{G\mathbb{R}}\{\mathbf{m}_{K+k} \otimes (x_{i,k} \oplus x_{j,k})\}\right\|_l. && (13)
\end{aligned}$$

The inequalities from (10) to (11) and from (12) to (13) stem from the properties of operations in the field of real numbers and GF, i.e.,

$$s_i - s_j \le 1_{G\mathbb{R}}(x_i \oplus x_j) \le s_i + s_j \tag{14}$$

where $x_i$ and $x_j$ are the GF representation of $s_i$ and $s_j$, respectively (see (1)). Moreover, $\mathbf{d} = [d(1) \ \cdots \ d(N-k)]^T$, where $d(k) = x_{i,k} \oplus x_{j,k}$, $0 \le i, j \le N$, as each element in $\mathbf{d}$ depends on two non-zero elements in each row of $\mathbf{D}$, and thus, on our choice of the additional constraints.

When the data $s_i$ and $s_j$ are very similar, the distance between them, $|s_i - s_j|$, becomes small, which leads to smaller values of $1_{G\mathbb{R}}(x_i \oplus x_j)$ with very high probability, i.e., the probability mass function is concentrated near $1_{G\mathbb{R}}(x_i \oplus x_j) = 0$ and it is decaying very sharply for larger $1_{G\mathbb{R}}(x_i \oplus x_j)$. If the data $s_i$ and $s_j$ have less similarity, however, it results in larger distance of $|s_i - s_j|$, leading to the probability mass function of $1_{G\mathbb{R}}(x_i \oplus x_j)$ is widely spread. This is shown experimentally in Appendix I. Therefore, given vectors $\mathbf{m}_{K+1}, \ldots, \mathbf{m}_N$ in (13), the error in the sense of similarity, i.e., $\|\mathbf{s} - \hat{\mathbf{s}}\|_l$, between the exact and approximate solutions decreases on average when the data have more similarity. ■

Property 1 implies that the decoding error is bounded, and that this bound becomes smaller when original data are more similar. This means that the best way to construct $\mathbf{D}$ consists in building additional constraints with source symbols that are expected to have the highest similarity. In order to show this analytically, consider $\mathbf{D}$ and $\tilde{\mathbf{D}}$ (with $\tilde{\mathbf{D}} \neq \mathbf{D}$), where $\tilde{\mathbf{D}}$ is constructed by a set of less similar data than $\mathbf{D}$ that is constructed by the most similar data. This means from (13) that the upper bounds of the

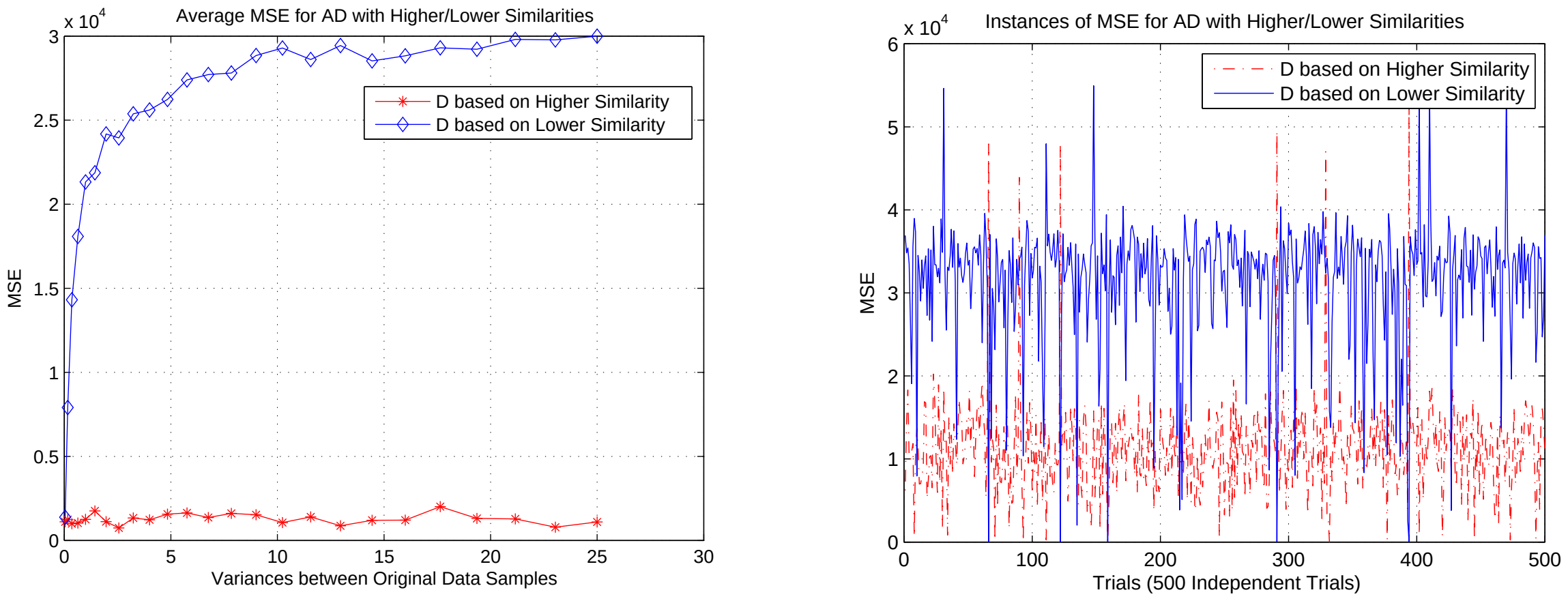

Fig. 2. Illustrative examples: performance comparison of the proposed approximate decoding algorithm with higher and lower similarities (i.e., with matrix $\mathbf{D}$ and $\tilde{\mathbf{D}}$). In order to emulate the higher and lower similarities in these examples, source data is generated as $s_i = s_1 + N(0, \sigma_i)$ where $s_1$ is given and $N(0, \sigma_i)$ denotes a zero mean Gaussian random variable with standard deviations $\sigma_2$ and $\sigma_3$. In these experiments, 1 of 3 packets is lost and approximate decoding algorithm uses $\mathbf{D} = [1\,1\,0]$ for higher similarity and $\mathbf{D} = [1\,0\,1]$ for lower similarity. Average performance (shown left with a fixed $\sigma_2 = 0.2$ and variable $\sigma_3 (\geq 0.2)$) and instantaneous performances ($\sigma_2 = 1, \sigma_3 = 10$) in independent experiments.

distance with $\mathbf{D}$ and $\tilde{\mathbf{D}}$ are respectively

$$\left\| \sum_{k=1}^{N-K} 1_{G\mathbb{R}} \left\{ \mathbf{m}_{K+k} \otimes (x_{i,k} \oplus x_{j,k}) \right\} \right\|_l \tag{15}$$

and

$$\left\| \sum_{k=1}^{N-K} 1_{G\mathbb{R}} \left\{ \mathbf{m}_{K+k} \otimes \{\tilde{x}_{i,k} \oplus \tilde{x}_{j,k}\} \right\} \right\|_l . \tag{16}$$

Since $x_{i,k}$ and $x_{j,k}$ are specified by $\mathbf{D}$ while $\tilde{x}_{i,k}$ and $\tilde{x}_{j,k}$ are specified by $\tilde{\mathbf{D}}$, it is true with high probability that

$$x_{i,k} \oplus x_{j,k} \leq \tilde{x}_{i,k} \oplus \tilde{x}_{j,k} \tag{17}$$

as discussed in Appendix I. Therefore, we can conclude from (9)–(13), (15), (16) and (17) that $\mathbf{D}$ leads to better performance (or equivalently less errors) than $\tilde{\mathbf{D}}$ on average if the approximate decoding is deployed in conjunction with the implementation proposed in Section II-C. An illustrative set of simulation results are shown in Fig. 2.

In summary, we observe that the efficiency of approximate decoding increases with the source similarity and with the accuracy about the correlation information that is used to derive additional constraints for decoding.

## IV. Optimal Finite Field Size

We study here the design of the coding coefficient matrix, and in particular, of the influence of the size of the finite field (i.e., GF) on the performance of the approximate decoding framework. This size has an influence on the reconstruction error when the number of symbols is insufficient for perfect decoding. The GF size determines the resolution of the source encoding since only a finite number of symbols (that is equal to the GF size) can be uniquely represented by the identity functions defined in Section II-A. Thus, as the GF size is enlarged, the error that may be incurred by quantizing source data becomes smaller. At the same time, however, there is higher probability that a large distortion is induced by the approximate reconstruction. We therefore determine the optimal GF size that minimizes the expected decoding error by trading off source approximation error and decoding error probability.

We first prove the following property, which states that the decoding errors increase as the GF size is enlarged. While this property seems contradictory, this is true because a perfect source model that identifies which source data are exactly the same is not available. Rather, the source model can only provide the information about the most similar data, so that the approximate decoding can use it for data recovery with best efforts. In the analysis, we consider a worse-case scenario, where data recovered by the constraints in the $\mathbf{D}$ matrix of the approximate decoding are uniformly distributed over $\mathcal{S}$.[4]

*Property 2: Given a finite set of data $\mathcal{S}$, the average reconstruction error increases as the GF size for the coding operations increases.*

*Proof:* Let $s \in \mathcal{S}$ be an original symbol, where the size of the original data space is given by $|\mathcal{S}| = 2^r$. Let further $\hat{s}_r = 1_{G\mathbb{R}}(\hat{x}_r)$ and $\hat{s}_R = 1_{G\mathbb{R}}(\hat{x}_R)$ be the decoded symbols when coding is performed in respectively GF($2^r$) and GF($2^R$) with $R > r$, for $r, R \in \mathbb{N}$, i.e., GF($2^R$) is an extended GF from GF($2^r$). In this scenario, the decoding errors are uniformly distributed over $\mathcal{S}$. Thus, the probability mass function of $\hat{s}_k$ is given by

$$p_k(\hat{s}_k) = \begin{cases} 1/2^k, & \text{if } \hat{s}_k \in [0, 2^k - 1] \\ 0, & \text{otherwise} \end{cases}$$

for $k \in \{r, R\}$. To prove that a larger GF size results in a higher decoding error, we have to show that

$$\Pr\left(|s - \hat{s}_R| \geq |s - \hat{s}_r|\right) > \frac{1}{2}. \tag{18}$$

If this condition is satisfied, the expected distortion is larger for $s_R$ than $s_r$, or equivalently, for the larger

[4]If distribution of the decoded data is known, it can be used for better approximate decoding. This may be an interesting future research direction.

GF size. The left hand side of (18) can be expressed as

$$\Pr\left(\hat{s}_R \geq \hat{s}_r, s \leq \frac{\hat{s}_R + \hat{s}_r}{2}\right) + \Pr\left(\hat{s}_R < \hat{s}_r, s > \frac{\hat{s}_R + \hat{s}_r}{2}\right)$$
$$= \Pr\left(\hat{s}_R \geq \hat{s}_r\right) \Pr\left(s \leq \frac{\hat{s}_R + \hat{s}_r}{2} \middle| \hat{s}_R \geq \hat{s}_r\right) + \Pr\left(\hat{s}_R < \hat{s}_r\right) \Pr\left(s > \frac{\hat{s}_R + \hat{s}_r}{2} \middle| \hat{s}_R < \hat{s}_r\right)$$
$$= \left(1 - 2^{r-R-1}\right)\hat{P} + 2^{r-R-1}\left(1 - \hat{P}\right) = 2^{r-R-1} + \left(1 - 2^{r-R}\right)\hat{P}$$

because $\hat{s}_R$ and $\hat{s}_r$ are both uniformly distributed. In the previous equations, we have posed $\hat{P} \triangleq \Pr\left(s \leq \frac{\hat{s}_R+\hat{s}_r}{2} \middle| \hat{s}_R \geq \hat{s}_r\right)$. We further show in Appendix II that $\hat{P} > \frac{1}{2}$. Therefore, we have

$$2^{r-R-1} + \left(1 - 2^{r-R}\right)\hat{P} > 2^{r-R-1} + \left(1 - 2^{r-R}\right) \cdot \frac{1}{2} = \frac{1}{2} \tag{19}$$

which completes the proof. ■

Property 2 implies that a small GF size is preferable in terms of expected decoding error. In particular, it is preferred not to enlarge the GF size more than the size of the input space since approximate decoding performs worse in very large field.

Alternatively, if the GF size becomes smaller than the size of the input alphabet size, the maximum number of source symbols that can be distinctively represented decreases correspondingly. Specifically, if we choose a GF size of $2^{r'}$ such that $|\mathcal{S}| > 2^{r'}$ for $r' < r$, part of the data in $\mathcal{S}$ needs to be discarded to form a subset $\mathcal{S}'$ such that $|\mathcal{S}'| \leq 2^{r'}$. In this case, we assume that if the GF size is reduced from GF($2^r$) to GF($2^{r-z}$), where $0 \leq z (\in \mathbb{Z}) \leq r-1$, the least significant $z$ bits in the representation of the original data are discarded first from $x \in \mathcal{S}$. Then, all the data in $\mathcal{S}'$ can be distinctly encoded in GF($2^{r'}$).

In summary, while reducing the GF size may result in lower decoding error, it may induce larger information loss in the source data. Based on this clear tradeoff, we propose Property 3 that shows the existence of an optimal GF size. Note that discarding part of source data information results in errors at the source, similar to data quantization. Thus, we assume that the corresponding source information loss is uniformly distributed and that the decoded data is also uniformly distributed in the following analysis.

*Property 3: There exists an optimal GF size that minimizes the expected error in data reconstruction at decoder. The optimal GF size is given by GF($2^{r-z^*}$), where $z^* = \lceil (r-1)/2 \rceil$ and $z^* = \lfloor (r-1)/2 \rfloor$.*

*Proof:* Suppose that the number of original source symbols is $|\mathcal{S}| = 2^r$ and that the coding field is GF($2^r$). As discussed in Property 2, the GF size does not need to be enlarged more than $2^r$, as this only increases the probability of the expected decoding error. If the GF size is reduced from GF($2^r$) to GF($2^{r-z}$), the approximate decoding is more efficient and the decoding errors are uniformly distributed over $[-r_D, r_D]$, where $r_D = 2^{r-1-z} - 1$, i.e.,

$$p_{e_D}(e_D) = \begin{cases} 1/(2r_D + 1), & \text{if } e_D \in [-r_D, r_D] \\ 0, & \text{otherwise} \end{cases}. \tag{20}$$

At the same time, if the GF size is reduced, the input data set $\mathcal{S}$ is reduced to $\mathcal{S}'$ and the number of input symbols is decreased. By discarding the $z$ least significant bits, the number of input symbols becomes $|\mathcal{S}'| = 2^{r-z}$. Such an information loss also results in errors over $[-r_I, r_I]$, where $r_I = 2^z - 1$, i.e.,

$$p_{e_I}(e_I) = \begin{cases} 1/(2r_I+1), & \text{if } e_I \in [-r_I, r_I] \\ 0, & \text{otherwise} \end{cases}. \tag{21}$$

Based on these independent distortions, the distribution of the total error, $p_{e_T}(e_T) = p_{e_D}(e_D) + p_{e_I}(e_I)$, is given by [24]

$$p_{e_T}(e_T) = \frac{H}{2}\left\{|e_T + r_I + r_D + 1| - |e_T + r_I - r_D| - |e_T - r_I + r_D| + |e_T - r_I - r_D - 1|\right\}$$

for $|e_T| \leq r_I + r_D \triangleq e_T^{max}$ and $H = (2r_I+1)^{-1}(2r_D+1)^{-1}$. Since $e_T + r_I + r_D + 1 \geq 0$ and $e_T - r_I - r_D - 1 \leq 0$ for all $|e_T| \leq e_T^{max}(= r_I + r_D)$, by substituting $r_I$ and $r_D$, we have

$$p_{e_T}(e_T) = \frac{H}{2}\left\{2\left(2^z + 2^{r-1-z} - 1\right) - \left|e_T + 2^z - 2^{r-1-z}\right| - \left|e_T - 2^z + 2^{r-1-z}\right|\right\}. \tag{22}$$

By denoting $a(z) \triangleq 2^z - 2^{r-1-z}$ and $b(z) \triangleq 2^z + 2^{r-1-z}$, the expected decoding error can be expressed as

$$E\left[|e_T|\right] = \sum_{e_T=-\infty}^{\infty} |e_T| \cdot p_{e_T}(e_T) = \sum_{e_T=-e_T^{max}}^{e_T^{max}} \frac{H}{2}|e_T| \cdot \left[2(b(z)-1) - |e_T + a(z)| - |e_T - a(z)|\right]. \tag{23}$$

Since both $|e_T|$ and $[2(b(z)-1) - |e_T + a(z)| - |e_T - a(z)|]$ are symmetric on $z = \lceil (r-1)/2 \rceil$ and $z = \lfloor (r-1)/2 \rfloor$ (see Appendix III), $E[|e_T|]$ is also symmetric. Thus,

$$\begin{aligned} E[|e_T|] &= H \sum_{e_T=1}^{e_T^{max}} e_T \cdot \{2(b(z)-1) - |e_T + a(z)| - |e_T - a(z)|\} \\ &= H \sum_{e_T=1}^{e_T^{max}} e_T \cdot \{2(b(z)-1)] - H \sum_{e_T=1}^{e_T^{max}} e_T \cdot \{|e_T + a(z)| + |e_T - a(z)|\} \\ &= H \cdot (b(z)-1)e_T^{max}(e_T^{max}+1) - H \sum_{e_T=1}^{e_T^{max}} e_T \cdot \{|e_T + a(z)| + |e_T - a(z)|\}. \end{aligned} \tag{24}$$

If we consider the case where $a(z) > 0$, which corresponds to $r/2 < z \leq r-1$, we have

$$\begin{aligned} \sum_{e_T=1}^{e_T^{max}} e_T \cdot \{|e_T + a(z)| + |e_T - a(z)|\} &= \sum_{e_T=1}^{a(z)-1} e_T \cdot 2a(z) + \sum_{e_T=a(z)}^{e_T^{max}} e_T \cdot 2e_T \\ &= \frac{1}{3}e_T^{max}(e_T^{max}+1)(2e_T^{max}+1) + \frac{1}{3}a(z)(a(z)^2-1). \end{aligned}$$

Note that $e_T^{max} = b(z) - 2$. Therefore, for the case where $a(z) > 0$, $E[e_T]$ can be expressed as

$$\begin{aligned} E[e_T] &= H \cdot \left[(b(z)-1)^2(b(z)-2) - \frac{1}{3}(b(z)-1)(b(z)-2)(2b(z)-3) - \frac{1}{3}a(z)(a(z)^2-1)\right] \\ &= H \cdot \left[\frac{1}{3}b(z)(b(z)-1)(b(z)-2) - \frac{1}{3}a(z)(a(z)^2-1)\right] \end{aligned} \tag{25}$$

which is an increasing function for $r/2 < z \leq r-1$ (see Appendix IV). Since $E[e_T]$ is symmetric on $z = \lceil (r-1)/2 \rceil$ and $z = \lfloor (r-1)/2 \rfloor$, and is an increasing function over $r/2 < z \leq r-1$, $E[e_T]$ is convex over $0 \leq z \leq r-1$. Therefore, there exists an optimal $z^*$ that minimizes the expected decoding error.

Finally, since $E[e_T]$ is symmetric on $\lceil (r-1)/2 \rceil$ and $\lfloor (r-1)/2 \rfloor$, the minimum $E[e_T]$ can be achieved if $z^* = \lceil (r-1)/2 \rceil$ and $z^* = \lfloor (r-1)/2 \rfloor$. The two optimum points can be the same for odd $r$. ■

## V. Approximate Decoding in Sensor Networks

### A. System Description

We illustrate in this section an example, where the approximate decoding framework is used to recover the data transmitted by sensors that capture a source signal from different spatial locations. We consider a sensor network, where sensors transmit RLNC encoded data. Specifically, each sensor measures its own observations and receives the other observations from its neighbor sensors. Then, each sensor combines the received data with its own data using RLNC. It transmits the resulting data to its neighbor nodes or receivers. In the considered scenario, there are 30 sensors which measure seismic signals placed at a distance of 100m by each other.

A sensor $h$ captures a signal $S_h$ that represents a series of sampled values in a time window of size $w$, i.e., $S_h = [s_h^1, \ldots, s_h^w]^T$. We assume that the data measured at each sensor are in the range of $[-s_{min}, s_{max}]$, i.e., $s_h^l \in \mathcal{S} = [-s_{min}, s_{max}]$ for all $1 \leq l \leq w$. We further assume that they are quantized and mapped to the nearest integer values, i.e., $s_h^l \in \mathbb{Z}$. Thus, if the measured data exceed the range of $[-s_{min}, s_{max}]$, then they are clipped to the minimum or maximum values of the range (i.e., $s_h^l = -s_{min}$ or $s_h^l = s_{max}$).

The data captured by the different sensors are correlated, as the signals at different neighboring positions are mostly time-shifted and energy-scaled versions of each other. The captured data have lower correlation with other signals as the distance between sensors becomes larger. An illustrative example is shown in Fig. 3(a) that presents seismic data recorded by 3 different sensors. The data measured by sensor 1 has much higher temporal correlation with the data measured by sensor 2 in terms of time shift and signal energy than the data measured by sensor 30. This is because sensor 2 is significantly closer to sensor 1 than sensor 30.

We consider that the nodes perform network coding for data delivery. We denote by $\mathbf{H}_n (\subseteq \mathbf{H})$ a set of sensors that are in the proximity of a sensor $n \in \mathbf{H}$. The number of sensors in $\mathbf{H}_n$ is $|\mathbf{H}_n| = N_n$. A sensor $n$ receives data $S_h$ from all the sensors $h \in \mathbf{S}_n$ in its proximity and encodes the received data with RLNC. The coding coefficients $c_h(k)$ are randomly selected from GF($2^r$) where the field size is determined such that $|\mathcal{S}| \leq 2^r$. The encoded symbols are then transmitted to the neighboring nodes or to the receiver. The

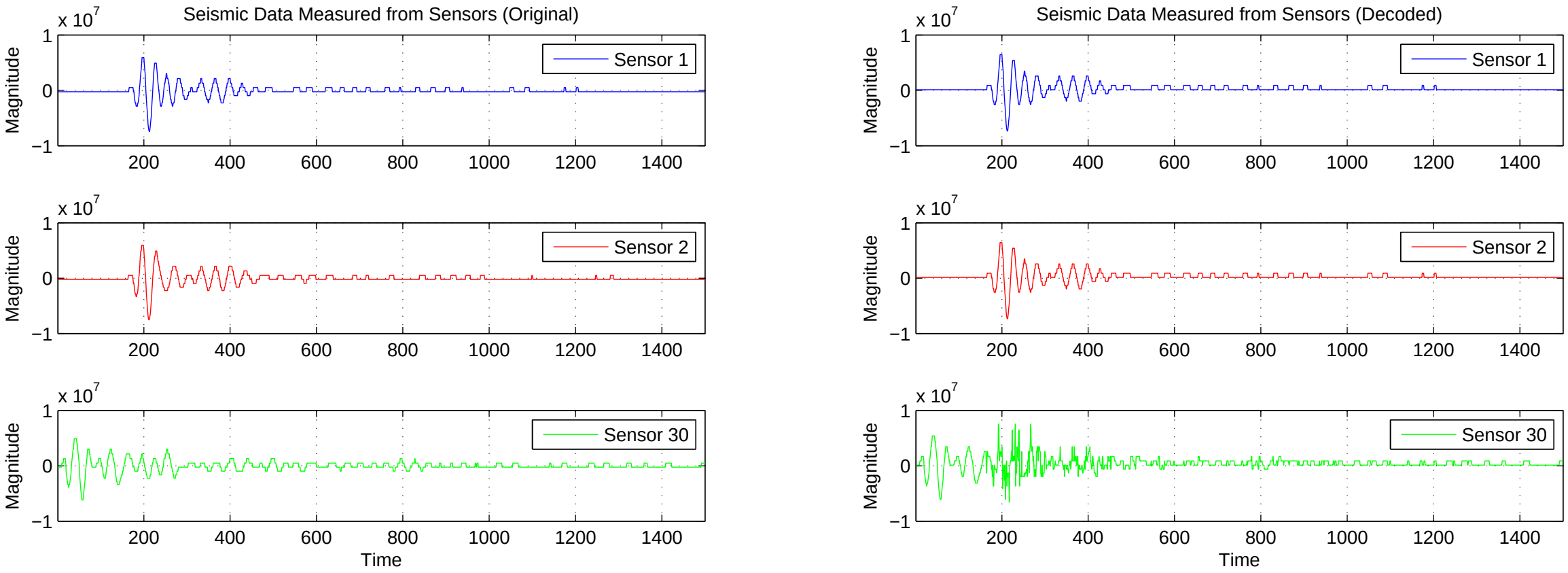

(a) (b)

Fig. 3. Measured original seismic data (a) and decoded seismic data based on approximate decoding (b).

$k$th encoded data packets for a window of samples are denoted by $Y(k) = \sum_{h \in \mathbf{H}_n} \bigoplus \{c_h(k) \otimes X_h\}$, where $X_h = 1_{\mathbb{R}G}(S_h)$. An illustrative example is shown in Fig. 4. This example presents a set of four sensors denoted by $\mathbf{H}$ that consists in two subsets of neighbors, i.e., $\mathbf{H}_1 = \{1, 3, 4\}$ and $\mathbf{H}_2 = \{2, 4\}$. The encoded data packets that the receiver collects from sensor 2 and sensor 4 are denoted by $Y(k_1)$ and $Y(k_2)$.

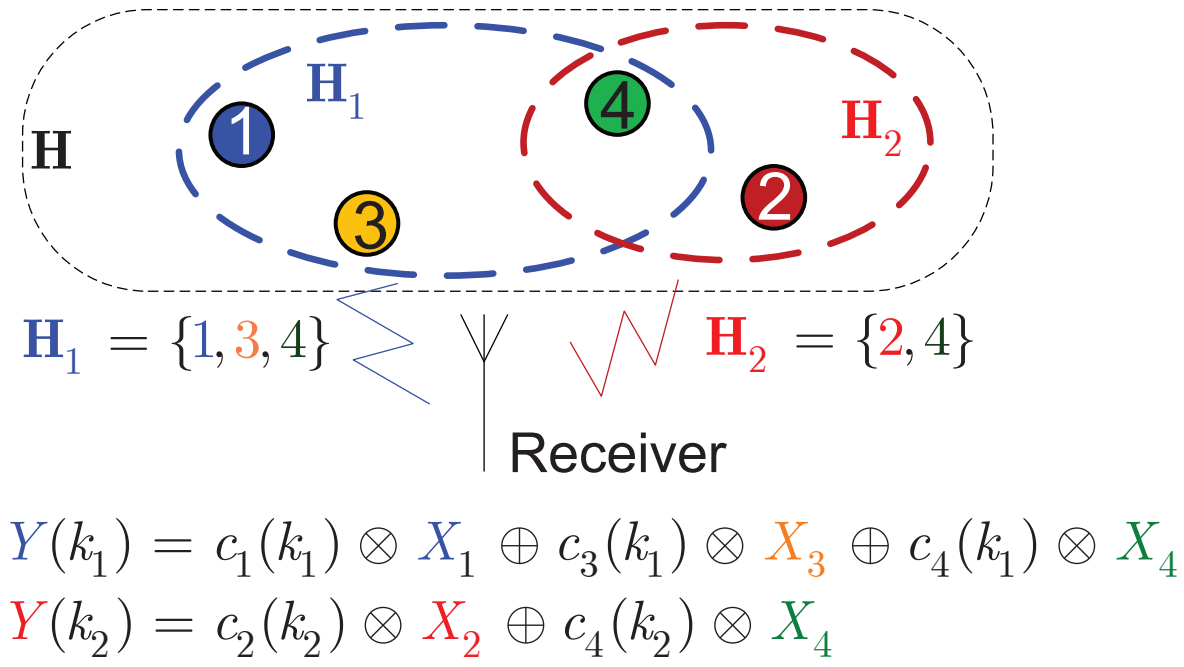

Fig. 4. Illustrative example of network coding in sensor networks.

When a receiver collects enough innovative packets, it can solve the linear system given in (6) and it can recover the original data. However, if the number of packets is not sufficient, the receiver applies our proposed approximate decoding strategy that exploits the similarity between the different signals. With such a strategy, the decoding performance can be improved as discussed in Property 1. We assume that the system setup is approximately known by the sensors. In other words, a simple correlation model can be computed, which includes the relative temporal shifts and energy scaling between the signals from the different sensors. In particular, since the sensor positions are known, one can simply assume that the data similarity depends only on the distance between sensors.

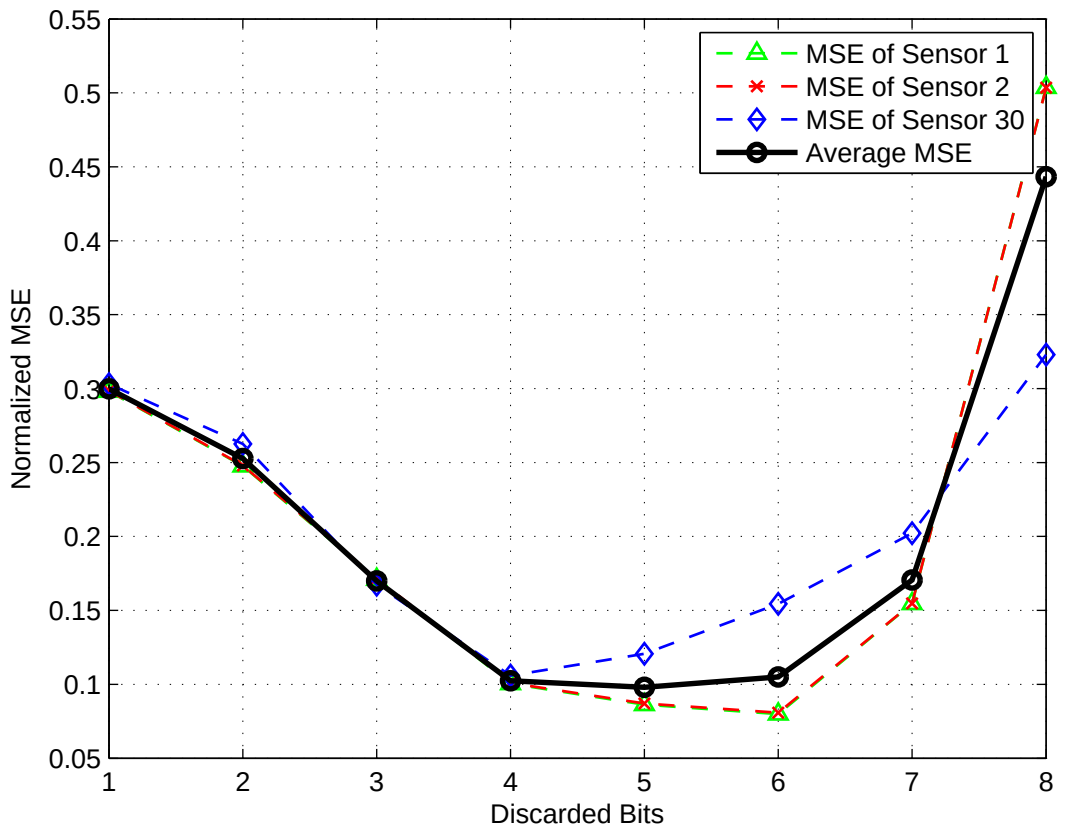

Fig. 5. Normalized average MSE for different GF sizes (i.e., GF($2^{10-z}$)).

## B. Simulation Results

We analyze an illustrative scenario, where the receiver collects encoded packets from sensors 1, 2 and 30 and tries to reconstruct the original signals from these three sensors. We consider temporal windows of size $w = 300$ for data representation. The captured data is in the range of $[0, 1023]$. Thus, the maximum GF size is $2^{10}$, i.e., GF($2^{10}$). We assume that $2/3$ of the linear equations required for perfect decoding are received with no error, and that the rest of $1/3$ of equations are not received. Thus, $1/3$ of the system constraints at decoder is built on, which is imposed into the coding coefficient matrix based on the assumption that the signals from sensor 1 and sensor 2 are highly correlated.

We study the influence of the size of the coding field on the decoding performance. Fig. 5 shows the MSE (Mean Square Error) distortion for the decoded signals for different number of discarded bits $z$, or equivalently for different GF sizes $2^{10-z}$. The conclusion drawn from Property 3 is confirmed from these results, as the decoding error is minimized at $z^* = \lceil (10-1)/2 \rceil = 5$.

An instantiation of seismic data recovered by the approximate decoding is further shown in Fig. 3, where a GF($2^{10-z^*}$) = GF($2^5$) is used. Since the additional constraints are imposed into the coding coefficient matrix based on the assumption of high correlation between the data measured by sensors 1 and 2, the recovered data of sensors 1 and 2 in Fig. 3(b) are very similar, but at the same time, the data are quite accurately recovered. We observe that the error in correlation estimation results in higher distortion in the signal recovered by sensor 30.

## VI. Approximate Decoding of Image Sequences

### *A. System Description*

In this section, we illustrate the application of approximate decoding to the recovery of image sequences. We consider a system, where information from successive frames is combined with network coding. Encoded packets are transmitted to a common receiver. Packets may, however, be lost or delayed, which prevents perfect reconstruction of the images. Thus, for improved decoding performance, we exploit the correlation between successive frames.

We consider a group of successive images in a video sequence. Each image $S_n$ is divided into $N$ patches $S_{n,p}$, i.e., $S_n = [S_{n,1}, \ldots, S_{n,N}]$. A patch $S_{n,p}$ contains $L \times L$ pixels $s^b_{n,p}$, $1 \le b \le L \times L$, i.e., $S_{n,p} = [s^1_{n,p}, \ldots, s^{L\times L}_{n,p}]$. Such a representation is illustrated on Fig. 6. The system implements RLNC and combines patches at similar positions in different frames to produce encoded symbols. In others words, it produces a series of symbols $Y_p(k) = \sum_{n=1}^{N} \bigoplus c_{n,p}(k) \otimes X_{n,p}$, where $X_{n,p} = 1_{\mathbb{R}G}(S_{n,p})$, for a location of patch $p$. The coding coefficients $c_{n,p}(k)$ are randomly chosen in $GF(2^r)$. We assume that the original data (i.e., pixels) can take values in $[0, 255]$, and thus, we choose the maximal size of the coding field to be $|\mathcal{S}| = 256 = 2^8$.

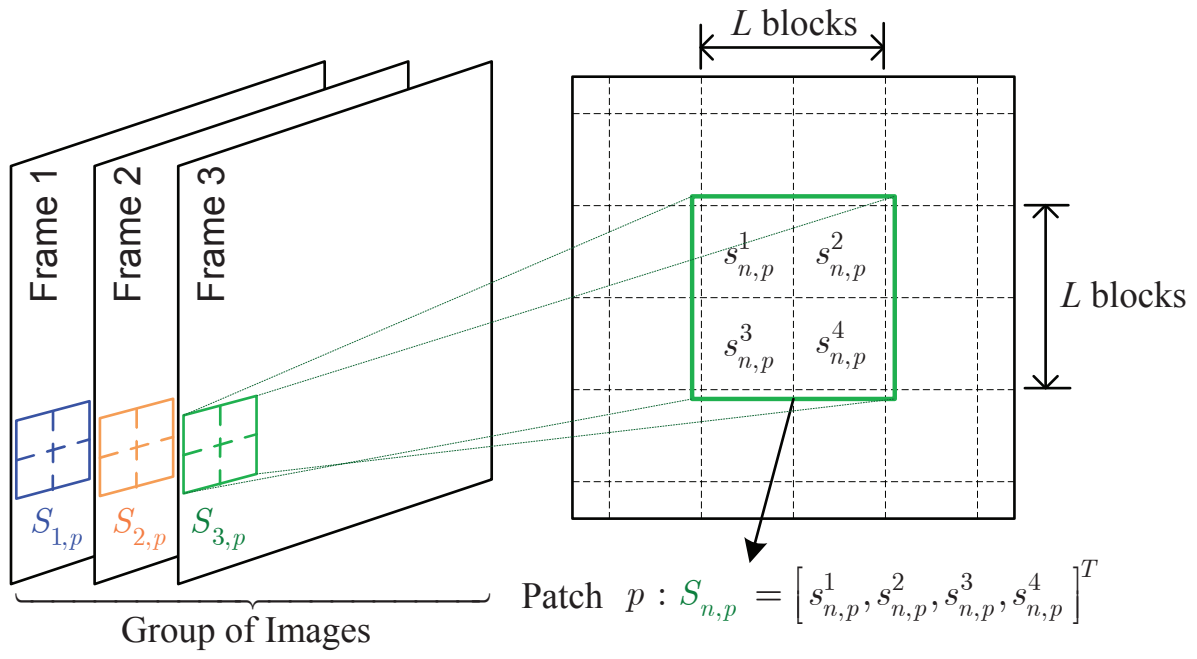

Fig. 6. Illustrative examples of patches in a group of images ($L = 2$).

When the receiver collects enough innovative symbols per patch, it can recover the corresponding sub-images in each patch, and eventually the group of images. If, however, the number of encoded symbols is insufficient, additional constraints are added to the decoding system in order to enable approximate decoding. These constraints typically depend on the correlation between the successive images. As an illustration, in our case, the constraints are imposed based on similarities between blocks of pixels in successive frames, i.e., $x^{b_1}_{n,p} = x^{b_2}_{n+1,p}$, where $1 \le b_1, b_2 \le L \times L$. The matched pixels, $b_1$ and $b_2$, are determined based on the motion information in successive image frames $n$ and $n + 1$, such that the similarity between patch $p$ is maximized. The motion information permits to add additional constraints to the decoding system so that estimations of the original blocks of data can be obtained by Gaussian elimination techniques. Due to our design choices, the decoding system can be decomposed into smaller independent sub-systems that correspond to patches.

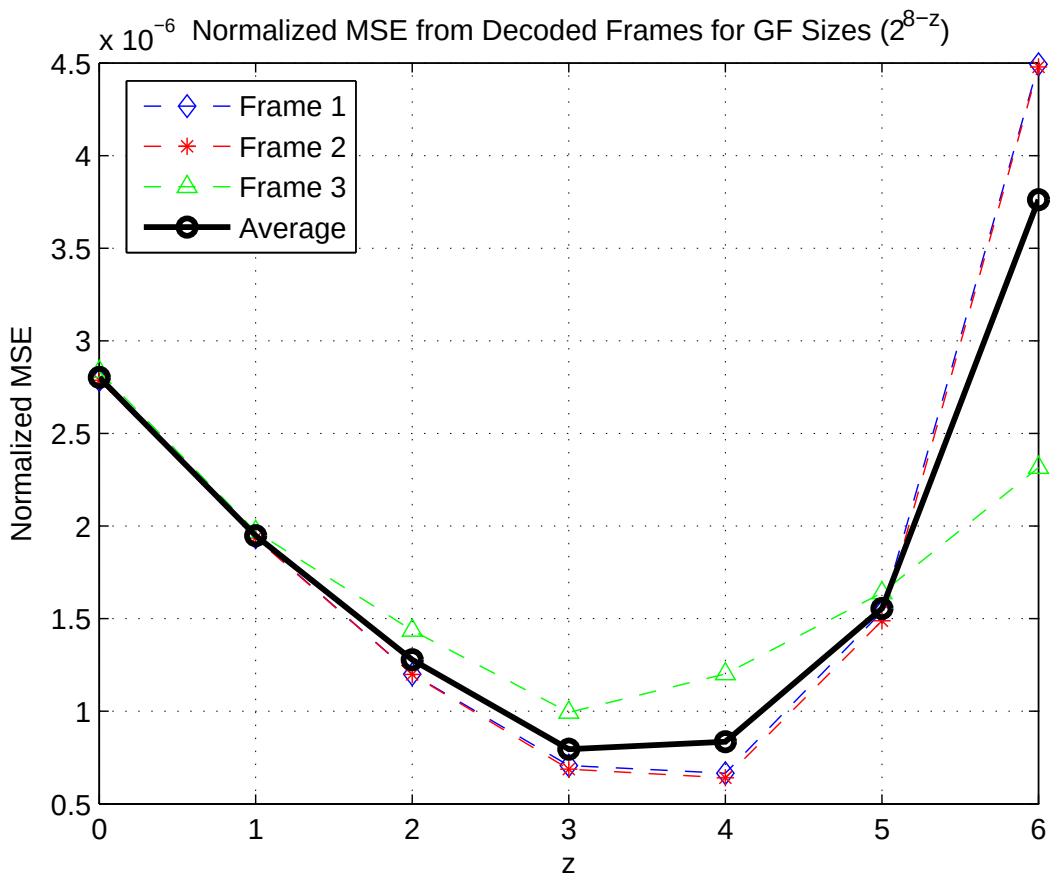

Fig. 7. Achieved Normalized MSE for different GF sizes (i.e., GF($2^{8-z}$)) in the approximate decoding of the *Silent* sequence.

## B. *Performance of Approximate Decoding*

In our experiments, we consider three consecutive frames extracted from the *Silent* standard MPEG sequence with QCIF format (174×144). The patches are constructed with four blocks of $8 \times 8$ pixels. We assume that only $2/3$ of the linear equations required for perfect decoding are received. The decoder implements approximate decoding by assuming that the correlation information is known at the decoder. The missing constraints are added to the decoding system based on the best matched pairs of blocks in consecutive frames, in the sense of the smallest distance (i.e, highest similarity) between the pixel values in blocks in different frames.

In the first set of experiments, we analyze the influence of the size of the coding field, by changing the GF sizes from GF($2^8$) to GF($2^{8-z}$). We reduce the size of the field by discarding the $z$ least significant bits for each pixel value. Fig. 7 shows the normalized MSE achieved from the decoded frames for different numbers of discarded bits $z$. As discussed in Property 3, the expected decoding error can be minimized if $z^* = \lceil (r-1)/2 \rceil$ and $z^* = \lfloor (r-1)/2 \rfloor$, which corresponds to $z^* = 3$ and $z^* = 4$. This can be verified from this illustrative example, where the maximum normalized MSE is achieved at $z = 4$ for frame 1 and frame 2, and at $z = 3$ for frame 3. The corresponding decoded images for two different GF sizes are presented in Fig. 8. From the decoded images, we can observe that several patches are completely black or white. This is because the coding coefficient matrices are singular, leading to the failure of Gaussian elimination during the decoding process. Note that the goal of results shown in Fig. 7 is to verify the Property 3, but is not to maximize the PSNR performance. In order to further improve the MSE performance, several image and video enhancement techniques such as error concealment [25] can be deployed.

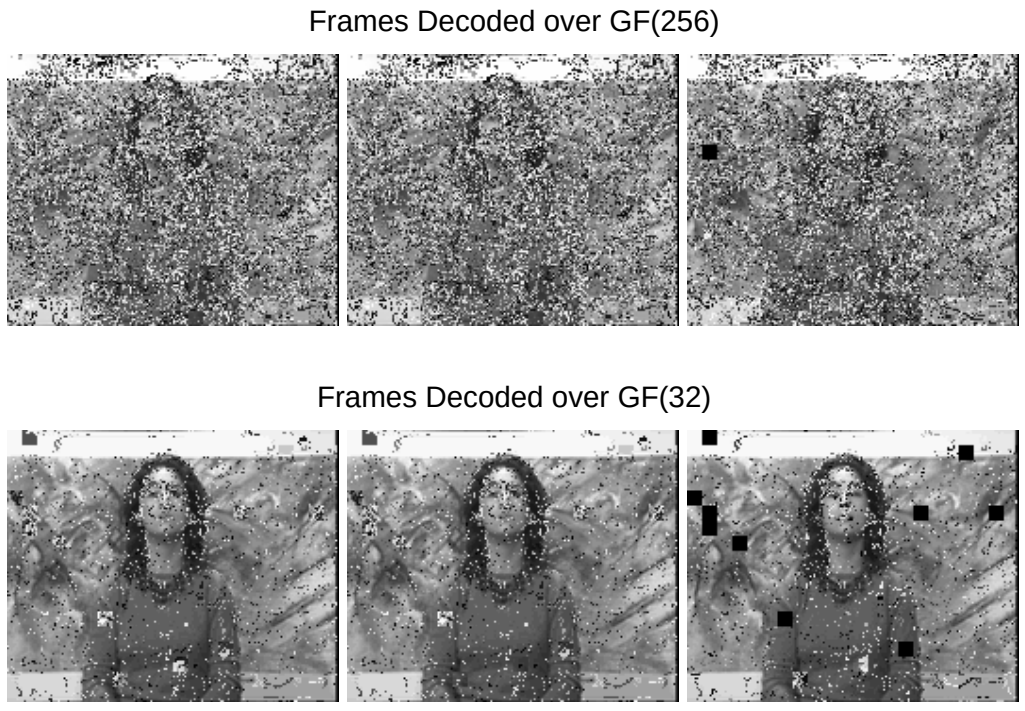

Fig. 8. Decoded frames for the *Silent* sequence for 2 different sizes of the coding field.

Next, we compare the approximate decoding approach with MLE based decoding for RLNC coded data, as the MLE can also use the joint probability distribution of sources for solving an undertermined system. In this experiment, our focus is on the case where clients receive a set of encoded packets that is insufficient for building a full-rank coefficient matrix, as this case is meaningful for both the approximate decoding and the MLE-based decoding. The source data are the first three frames of QCIF *Foreman* and Silent sequences. They have different characteristics as *Foreman* sequence has much higher motion than *Silent* sequence. For fair comparison, the same correlation information, i.e., the most similar data should be set equal, is used both for the MLE decoding and approximate decoding. For the approximate decoding, we assume that if the Gaussian elimination for a patch fails due to the singular coefficient matrix having $\mathbf{D}$ constraints, the resulting decoded patch is set to the average value of image pixel blocks. This choice is motivated by the fact that the MLE-based decoding always selects a solution even though the selected solution is not the best.

The results are presented in Fig. 9 with respect to the number of discarded bitplanes $z$, where the size of GF is determined by $GF(2^{8-z})$. From Fig. 9(a), we can observe that the approximate decoding outperforms the MLE for Silent sequence in all range of $z$ values. While the MLE shows a better performance than the approximate decoding for Foreman sequence in Fig. 9(b), there are several values of the GF sizes that show similar performance for both methods. The gain of the MLE for Foreman sequence mainly comes from the selection of brighter colors for representing the blocks, while the approximate decoding selects grayer colors.

However, in terms of complexity, the approximate decoding requires significantly less complexity than the MLE, as the Gaussian elimination is applied to very sparse matrices. In particular, assume that we have $y$ unknowns and $x$ equations are received. Then, it is known that the Gaussian elimination requires asymptotically at most $O(y^3)$ operations [26], while the MLE with exhaustive search requires asymptotically $O(q^{y-x}x^3)$, where $q \geq 2$ is a GF size [20] (Part II). As $y$ increases, $q^{y-x}$ increases much faster than $(y/x)^3$, which means that the approximate decoding can perform significantly faster than MLE-based approach. Therefore, we can conclude that the approximate decoding represents an effective solution for decoding with insufficient data and moderate complexity.

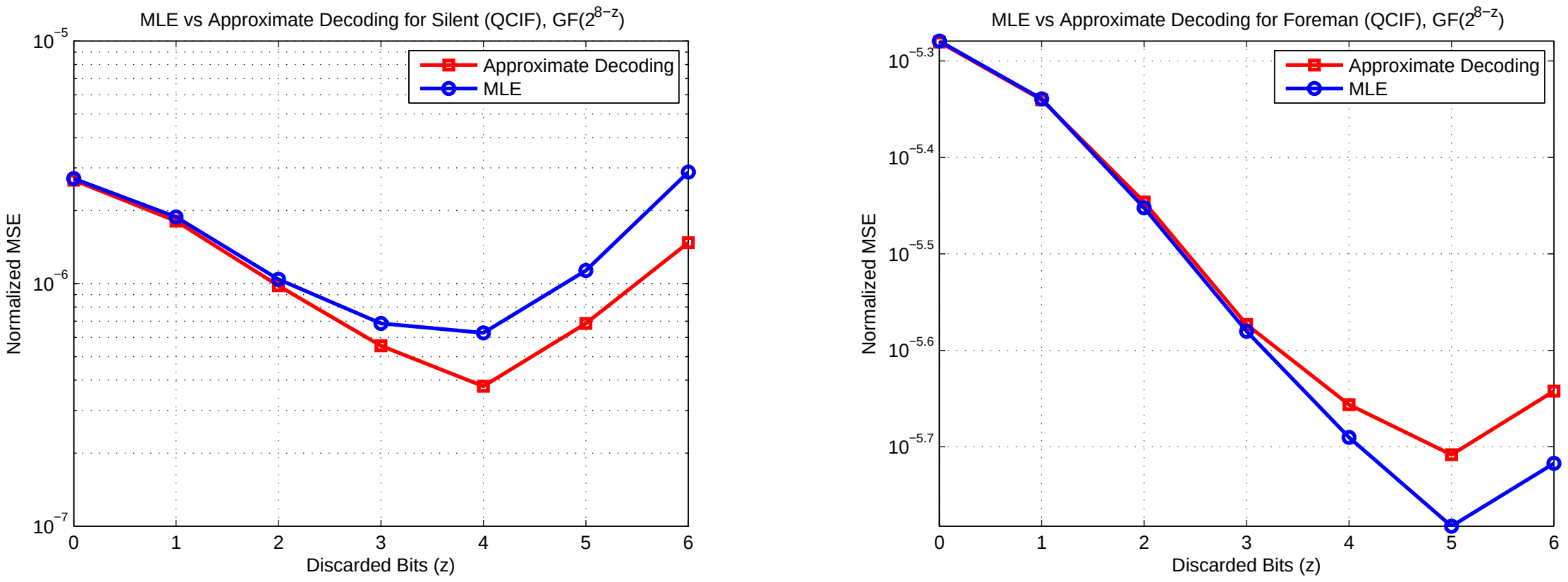

(a) (b)

Fig. 9. Performance comparison of the proposed approximate decoding method with MLE based decoding with respect to the number of discarded bitplanes $z$ for : (a) Silent QCIF and (b) Foreman QCIF sequence.

We also illustrate the influence of the accuracy of the correlation information by considering zero motion at the decoder. In other words, additional constraints for approximate decoding simply impose that the consecutive frames are identical. Fig. 10 shows the frames decoded with no motion over GF(32). We can see that the first three frames still provides an acceptable quality since the motion between these frames is actually very small. However, in frames 208, 209, and 210, where motion is higher, we clearly observe significant performance degradation, especially in the positions where high motion exists.

Next, we study the influence of the size of the group of images (i.e., window size) that is considered for encoding. It has been discussed that the coding coefficient matrices can be singular, as the coefficients are randomly selected in a finite field. This results in performance degradation for the approximate decoding. Moreover, it is shown that the probability that random matrices over finite fields are singular becomes smaller as the size of matrices becomes larger [27]. Thus, if the group of images (i.e., window size) becomes larger, the coding coefficient matrix becomes larger. As a result, the probability that Gaussian elimination fails is correspondingly smaller. This is quantitatively investigated from the following

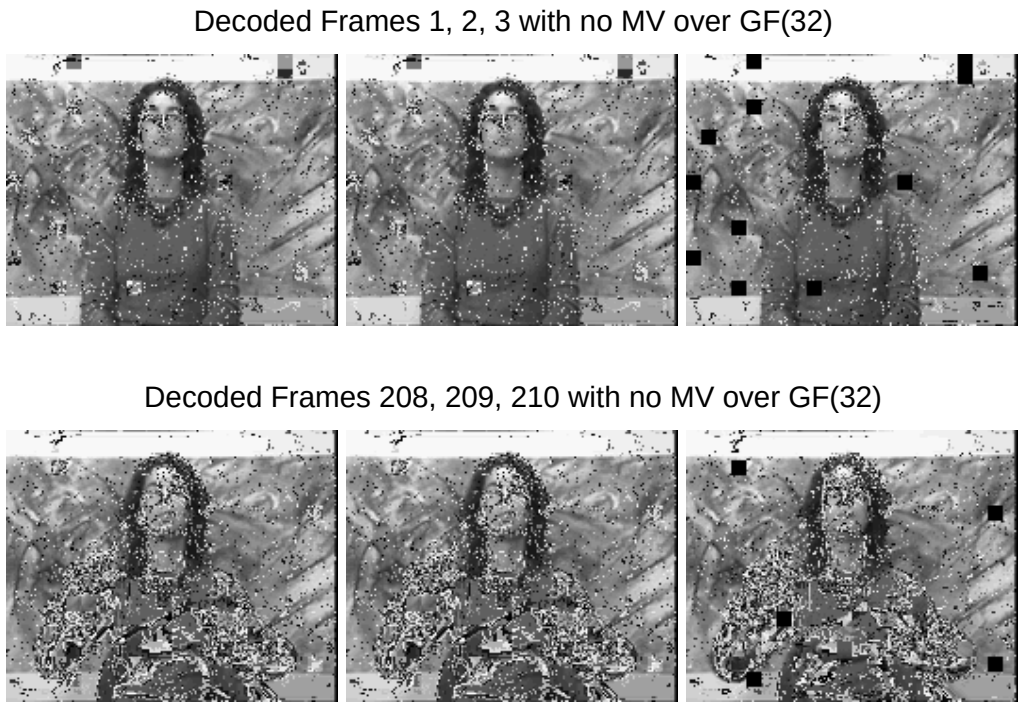

Fig. 10. Decoded frames with no information about motion estimation.

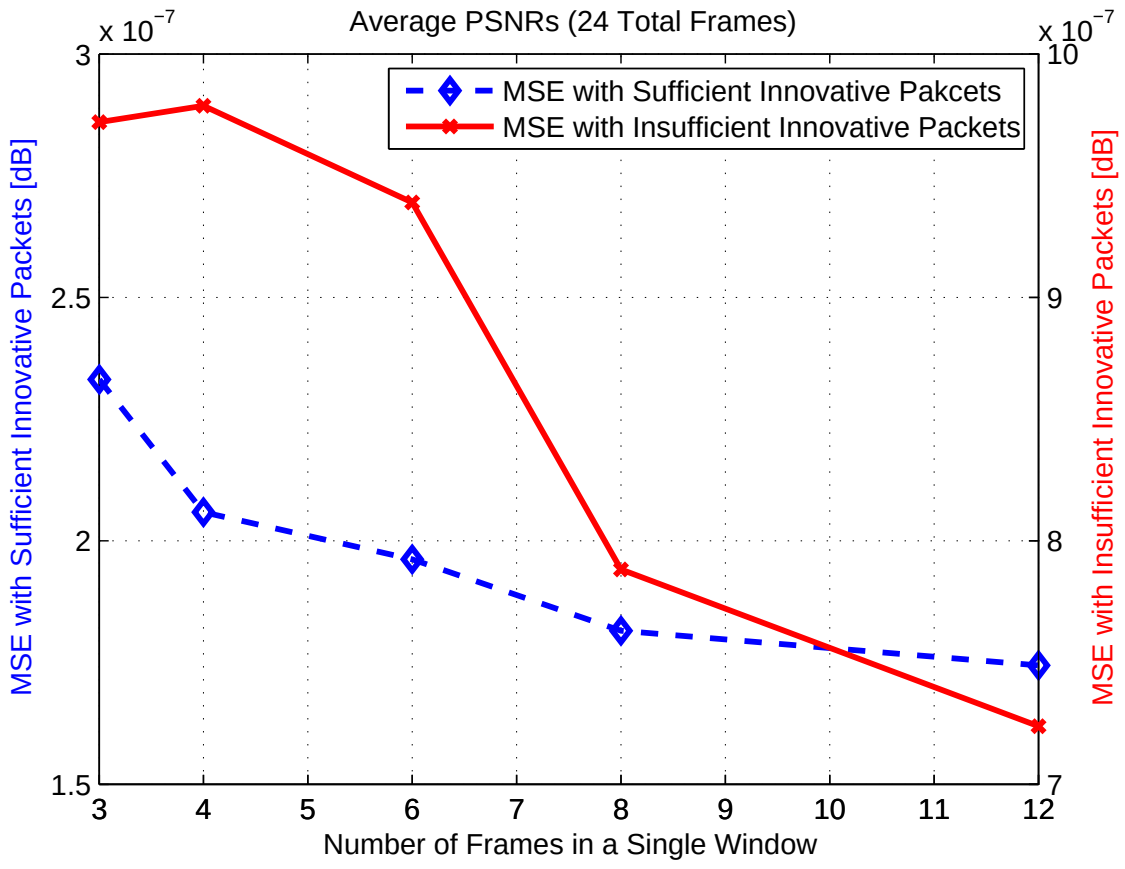

Fig. 11. Decoding MSE for different window sizes in the encoding of the sequence *Silent*.

experiment.

We consider 24 frames extracted from the *Silent* sequence and a set of different window sizes that contains 3, 4, 6, 8, and 12 frames. For example, if window size is 3, then there are 24/3=8 windows that are used in this experiment. The average normalized MSE achieved in the lossless case, where the decoder receives enough packets for decoding, is presented in Fig. 11. The normalized MSE decreases as the window sizes are enlarged. The only reason why all the frames are not perfectly recovered is the failure of the Gaussian elimination, when the coding coefficient matrices becomes singular. This confirms the above-mentioned discussion, i.e., if window size increases, the size of coding coefficient matrix also increases. Since the probability that the enlarged coding coefficient matrices are singular becomes smaller, higher average MSEs can correspondingly be achieved for larger size of window.

Finally, we study the influence of the window size in the lossy case. We assume that we have a loss rate of $1/24$ in all the configurations and the approximate decoding is implemented. Fig. 11 shows the achieved average MSE across the recovered frames for different window sizes. Since the decoding errors incurred by the approximate decoding are limited to a window and do not influence the decoding of the other windows, a small window size is desirable for limited error propagation. However, as discussed, a smaller window size can result in higher probability that the coding coefficient matrices become singular, and that the Gaussian elimination fails. Due to this tradeoff, we can observe that the achieved MSE becomes high when window size is 4 in our example. Note that the computational complexity for decoding (i.e., Gaussian elimination) also increases as the window size increases. Hence, the proper window size needs to be determined based on several design tradeoffs in practice.

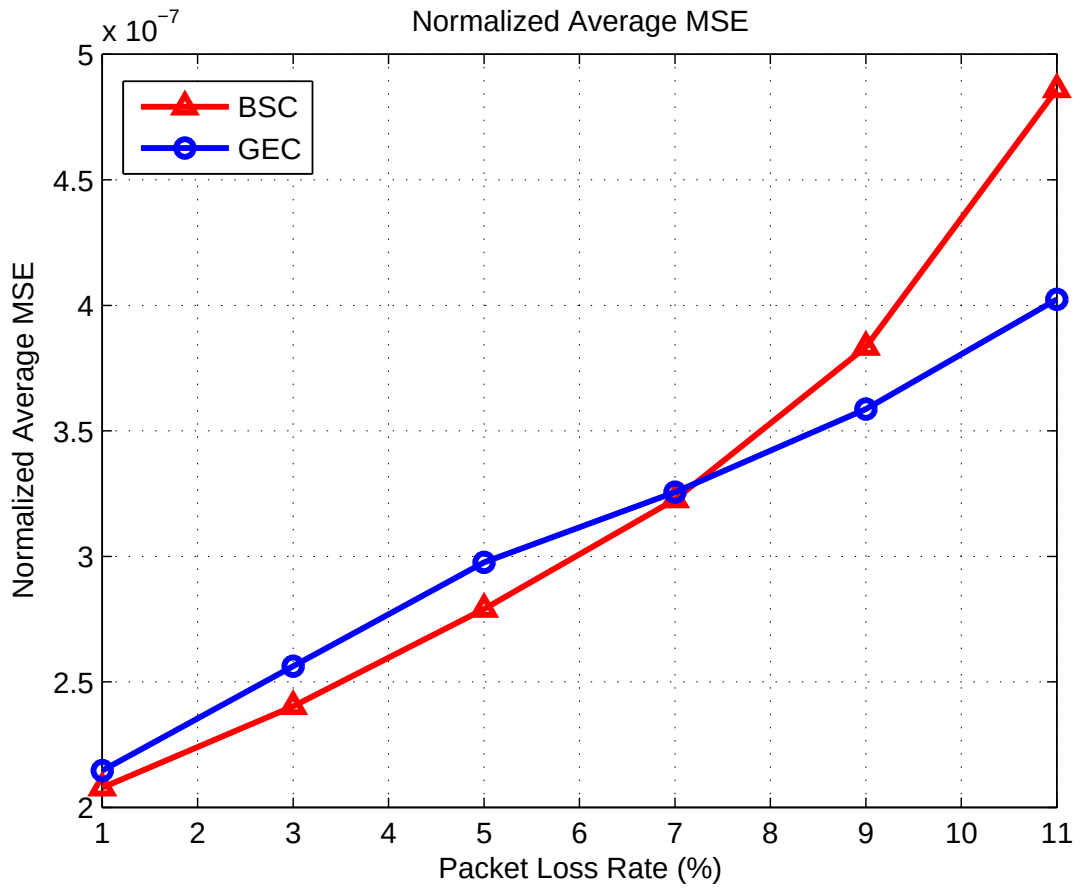

Fig. 12. Average MSE for the transmission of *Container* QCIF sequence with respect to various packet loss rates in the BSC and in GEC. Network coding is performed on windows of four frames.

### *C. Performance in Various Network Conditions*

We thus far considered a network having a fixed packet loss rate (i.e., a dedicated final node receives 2/3 of the required linear equations and does not receive 1/3 of the required linear equations). We now examine more general network scenarios, which may result in different packet loss rates for the final decoder. As an illustration, we consider a network which consists of three pairs of sources and destinations with several network nodes performing network coding operations. We assume that there are no loss in the sources and destinations and they are properly dimensioned. However, the links between nodes performing network coding operations are lossy with different packet loss rates. We study the achieved performance (MSE) that corresponds to different packet loss rates. The results are shown in Fig. 12. These results show the average MSE that the final node achieves when it experiences a variety of packet loss rates and decodes the received data with the proposed approximate decoding method for binary symmetric channel (BSC) and Gilbert Elliot channels (GEC) [28], respectively. The source images are from *Container* sample MPEG sequences with QCIF resolution. In all cases, the data is encoded with RLNC and a window of four packets is considered. We simulate loss with a GEC model [28] that consists in a two-state Markov chain where the good and bad states represent the correct reception or the loss of a packet, respectively. We choose the average length of burst of errors to 9 packets, and we vary the average packet loss rate in order to study the performance of our approximate reconstruction algorithm in different channel conditions. For the BSC model, the experiments are performed with a set of different average packet loss rates. As expected, the performance worsens as the packet loss rate increases. Moreover, these results show that the approximate decoding enables the decoder to achieve a noticeable gain in terms of decoded quality compared to traditional network coding based systems, which may completely fail to recover data. Alternatively, this means that the approximate decoding may require less network loads than traditional decoding algorithms in order to achieve the same decoding quality.

## VII. Conclusions

In this paper, we have described a framework for the delivery of correlated information sources with help of network coding along with a novel low complexity approximate decoding algorithm. The approximate decoding algorithm permits to reconstruct an approximation of the source signals even when an insufficient number of innovative packets are available for perfect decoding. We have analyzed the tradeoffs between the decoding performance and the size of the coding fields. We have determined an optimal field size that leads to the highest approximate decoding performance. We also have investigated the impact of the accuracy of the data similarity information used in building the approximate decoding solution. The proposed approach is implemented in illustrative examples of sensor network and distributed imaging applications, where the experimental results confirm our analytical study as well as the benefits of approximate decoding solutions as an efficient way to decode undertermined systems with reasonable complexity when source data are highly correlated.

## VIII. Acknowledgments

The authors would like to thank Dr. Laurent Duval for providing the seismic data used in the sensor network example.

## Appendix I

In this appendix, we provide illustrative examples that verify the arguments, where smaller values of $|s_i - s_j|$ can indeed lead to smaller values of $1_{G\mathbb{R}}(x_i \oplus x_j)$, which is discussed in the proof of Property 1. In this example, we consider GF(512), and study several examples of $|s_i - s_j| = 0, 1, 2, 50, 100, 150, 256$.

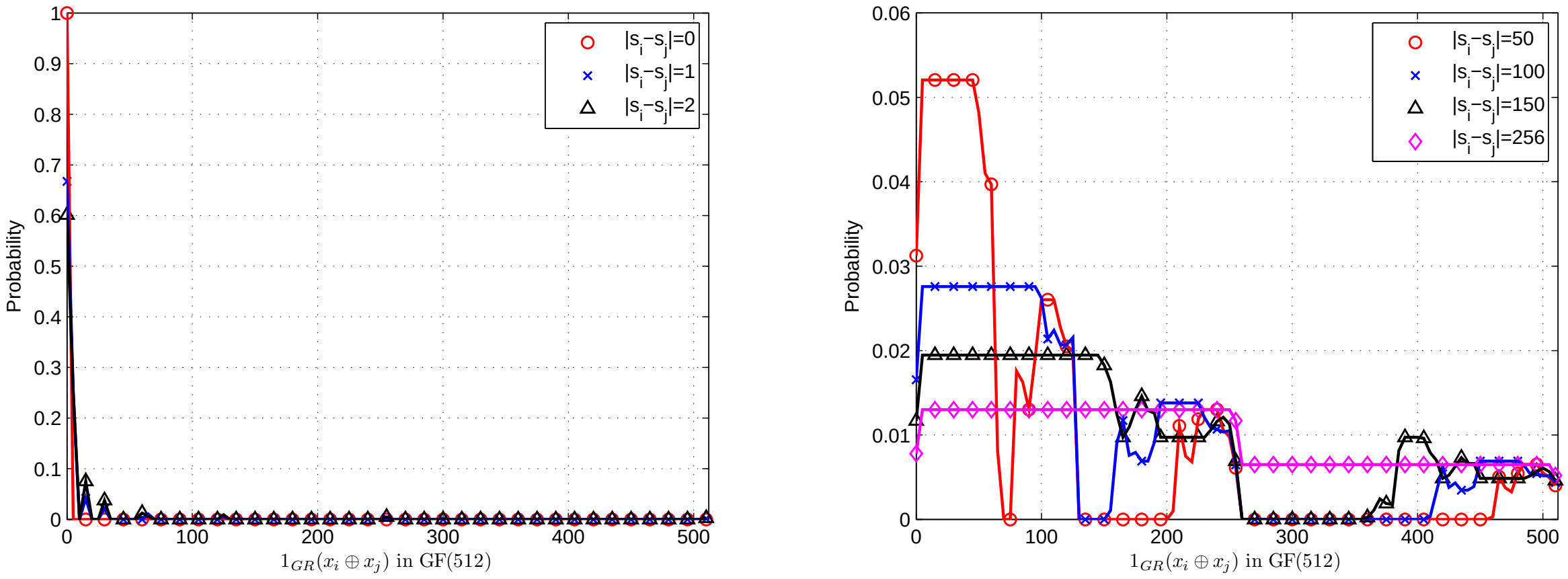

(a) (b)

Fig. 13. The probability mass function for different values of $1_{G\mathbb{R}}\left(x_i \oplus x_j\right)$ corresponding to various values of $|s_i - s_j|$ in GF(512).

In the cases where smaller differences between $s_i$ and $s_j$ (e.g., $|x_i - x_j| = 0, 1, 2$), we can observe that the most of the values of $1_{G\mathbb{R}}\left(x_i \oplus x_j\right)$ are concentrated around 0. In the cases where larger differences between $s_i$ and $s_j$ (e.g., $|s_i - s_j| = 50, 100, 150, 256$), however, the values of $1_{G\mathbb{R}}\left(x_i \oplus x_j\right)$ are spread over the elements in GF. Therefore, it is obviously confirmed that smaller values of $|s_i - s_j|$ indeed result in $1_{G\mathbb{R}}\left(x_i \oplus x_j\right)$. These are depicted in Fig. 13.

## Appendix II

In this appendix, we show that $\hat{P} \geq \frac{1}{2}$, where $\hat{P}$ is defined as $\hat{P} \triangleq \Pr\left(s \leq \frac{\hat{s}_R + \hat{s}_r}{2} \middle| \hat{s}_R \geq \hat{s}_r\right)$ in (19). Note that both $\hat{s}_r$ and $\hat{s}_R$ are reconstructed data, and thus, they are real values. Using Bayes' rule,

$$\begin{aligned}\hat{P} = \Pr\left(s \leq \frac{\hat{s}_R + \hat{s}_r}{2} \middle| \hat{s}_R \geq \hat{s}_r\right) &= \sum_{z=0}^{2^r-1} \Pr\left(z \leq \frac{\hat{s}_R + \hat{s}_r}{2} \middle| \hat{s}_R \geq \hat{s}_r, s = z\right) \Pr\left(s = z\right) \\ &= \frac{1}{2^r} \sum_{z=0}^{2^r-1} \Pr\left(z \leq \frac{\hat{s}_R + \hat{s}_r}{2} \middle| \hat{s}_R \geq \hat{s}_r, s = z\right).\end{aligned}$$

Referring to Fig. 14, we have

$$\begin{aligned}\sum_{z=0}^{2^r-1} \Pr\left(z \leq \frac{\hat{s}_R + \hat{s}_r}{2} \middle| \hat{s}_R \geq \hat{s}_r, s = z\right) &= \frac{1}{2^{r+R}} \sum_{z=0}^{2^r-1} \left[2^{r+R} - \left\{2^{r-1}(2^r - 1) + 2\sum_{l=0}^{z} l\right\}\right] \\ &= \frac{1}{2^{r+R}} \left\{2^{2r+R} - \frac{1}{6}\left(5 \cdot 2^{3r} - 3 \cdot 2^{2r} - 2 \cdot 2^r\right)\right\}.\end{aligned}$$

Thus, $\hat{P}$ can be expressed as

$$\hat{P} = \frac{1}{2^r}\left[\frac{1}{2^{r+R}}\left\{2^{2r+R} - \frac{1}{6}\left(5 \cdot 2^{3r} - 3 \cdot 2^{2r} - 2 \cdot 2^r\right)\right\}\right] = 1 - \frac{1}{6}\left[5 \cdot \frac{2^r}{2^R} - \frac{3}{2^R} - \frac{2}{2^{r+R}}\right].$$

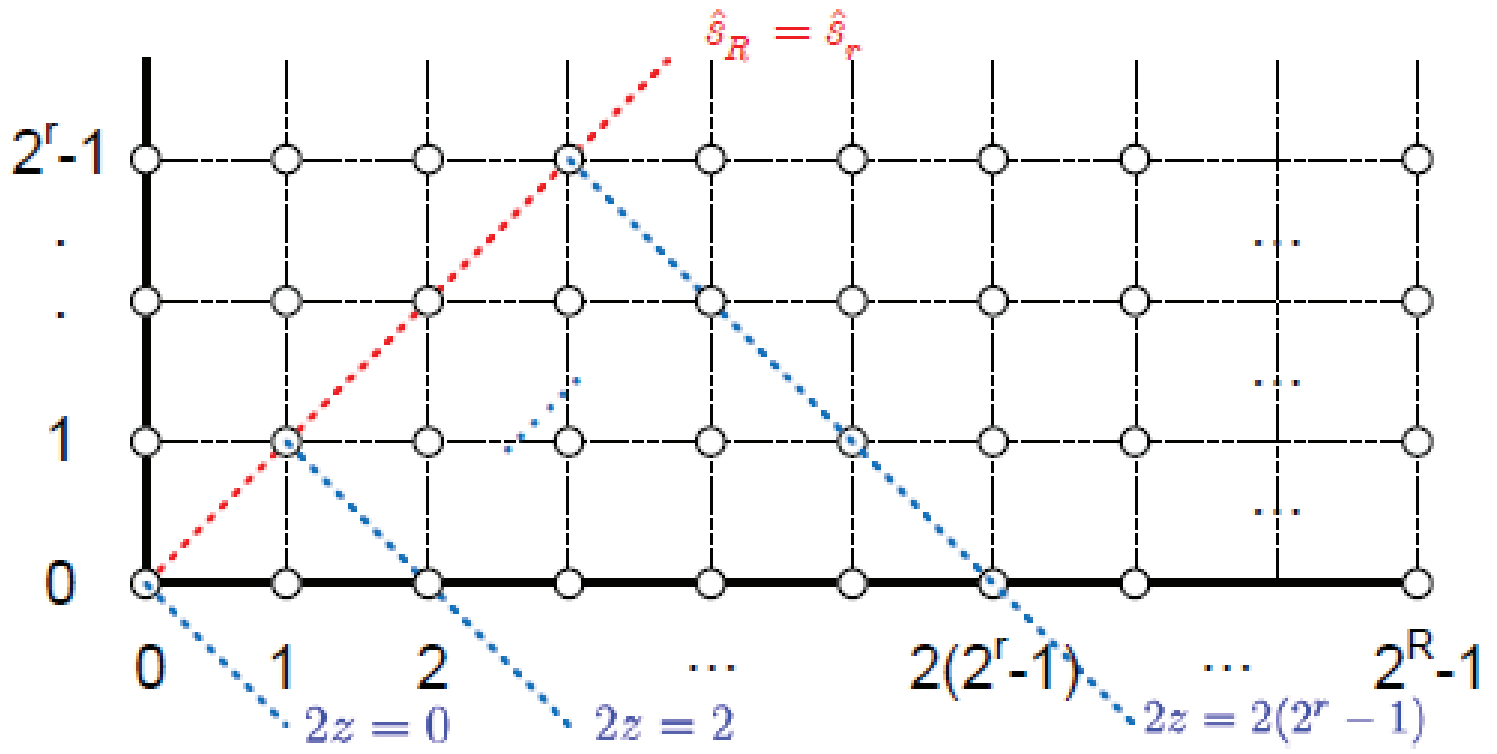

Fig. 14. An illustration for Appendix II

Since $r, R \in \mathbb{N}$ and $R > r$, $R$ can be expressed as $R = r + \alpha$, where $\alpha \in \mathbb{N}$. Thus,

$$\hat{P} = 1 - \frac{1}{6}\left[5 \cdot \frac{1}{2^{\alpha}} - \frac{3}{2^{r+\alpha}} - \frac{2}{2^{2r+\alpha}}\right].$$

Since $\lim_{r\to\infty} \hat{P} = 1 - \frac{5}{6} \cdot \frac{1}{2^{\alpha}} > \frac{1}{2}$ for all $\alpha \in \mathbb{N}$ and $\hat{P}$ is a non-increasing function of $r$, $\hat{P} > \frac{1}{2}$ for all $r, R$.

## Appendix III

In this appendix, we prove that the function $g(z) = 2(b(z)-1) - |e_T + a(z)| - |e_T - a(z)|$ is symmetric on $\lceil (r-1)/2 \rceil$, which is used in the proof of Property 3. To show this, we need to prove that $g(z) = g(r-1-z)$ for all $0 \le z(\in \mathbb{Z}) \le r-1$. Note that $a(r-1-z) = 2^{r-1-z} - 2^{r-1-(r-1-z)} = -(2^z - 2^{r-1-z}) = -a(z)$ and $b(r-1-z) = 2^{r-1-z} + 2^{r-1-(r-1-z)} = 2^z + 2^{r-1-z} = b(z)$. Thus,

$$g(r-1-z) = 2(b(r-1-z)-1) - |e_T + a(r-1-z)| - |e_T - a(r-1-z)|$$

$$= 2(b(z)-1) - |e_T - a(z)| - |e_T + a(z)| = g(z)$$

which completes the proof.

## Appendix IV

In this appendix, we show that

$$h(z) = \frac{1}{3}b(z)(b(z)-1)(b(z)-2) - \frac{1}{3}a(z)(a(z)^2 - 1) \quad (26)$$

is an increasing function for $z \in \mathbb{Z}$ where $r/2 < z \le r-1$. This is used in the proof of Property 3 Note that (26) is equivalent to function $h(z)$ with $z \in \mathbb{R}$ where $r/2 < z \le r-1$, sampled at every $z \in \mathbb{Z}$.

Thus, we focus on showing that $h(z)$ is an increasing function over $z \in \mathbb{R}$ where $r/2 < z \leq r-1$. To show that $h(z)$ is an increasing function, we may show that $\frac{d}{dz}h(z) > 0$ for $r/2 < z \leq r-1$. Note that

$$\frac{d}{dz}a(z) = \ln 2 \cdot (2^z + 2^{r-1-z}) = b(z)\ln 2$$

and

$$\frac{d}{dz}b(z) = \ln 2 \cdot (2^z - 2^{r-1-z}) = a(z)\ln 2.$$

Therefore,

$$\begin{aligned}\frac{d}{dz}h(z) &= \frac{\ln 2}{3}\left\{\left(3b(z)^2\frac{db(z)}{dz} - 6b(z)\frac{db(z)}{dz} + 2\frac{db(z)}{dz}\right) - \left(3a(z)^2\frac{da(z)}{dz} - \frac{da(z)}{dz}\right)\right\} \\ &= \frac{\ln 2}{3}\left\{3a(z)b(z)(b(z)-a(z)-2) + 2a(z) + b(z)\right\}.\end{aligned}$$

Since $a(z)b(z) = 2^{2z} - 2^{2(r-1-z)} > 0$ and $b(z) - a(z) = 2 \cdot 2^{r-1-z} \geq 2$ for $r/2 < z \leq r-1$,

$$\frac{d}{dz}h(z) = \frac{\ln 2}{3}\left\{3a(z)b(z)(b(z)-a(z)-2) + 2a(z) + b(z)\right\} > 0$$

which implies that $h(z)$ is an increasing function over $r/2 < z \leq r-1$.